\RequirePackage{lineno}
\documentclass[prc,preprint,superscriptaddress,showpacs,amssymb,amsmath,amsfonts,aps,floatfix]{revtex4}
\usepackage{graphicx}
\usepackage{dcolumn}
\usepackage{bm}
\usepackage{epsfig}
  
\def\be{\begin{equation}}  
\def\ee{\end{equation}}  
\def\bea{\begin{eqnarray}}  
\def\eea{\end{eqnarray}}  
\def\tpII{$e p \rightarrow e \gamma\gamma p$ }
\def\tpIIn{$e p \rightarrow e \gamma\gamma p$}
\def\tpIIr{$e p \rightarrow e \pi^0 p$ }

\def\tpVI{$e p \rightarrow e \gamma(\gamma) p$ }
\def\tpVIn{$e p \rightarrow e \gamma(\gamma) p$}
\def\tpIX{$e p \rightarrow e \gamma\gamma (p)$ }
\def\tpIXn{$e p \rightarrow e \gamma\gamma (p)$}
\def\mxepi{$M_x^{e\pi}$ }

\def\mxepg{M_x^{ep\gamma}}
\def\cthcm{$\cos(\theta^*)$}
\def\cthcmsp{$\cos(\theta^*)$ }

\def\phicm{$\phi^*$}
\def\phicmsp{$\phi^*$ }

\def\>{\hskip .1in }

\begin{document}

\title{Target and beam-target spin asymmetries in exclusive 
pion electroproduction for $Q^2>1$ GeV$^2$. II. $e p \rightarrow e \pi^0 p$} 

\newcommand*{\ANL}{Argonne National Laboratory, Argonne, Illinois 60439}
\newcommand*{\ANLindex}{1}
\affiliation{\ANL}
\newcommand*{\ASU}{Arizona State University, Tempe, Arizona 85287-1504}
\newcommand*{\ASUindex}{2}
\affiliation{\ASU}
\newcommand*{\CSUDH}{California State University, Dominguez Hills, Carson, CA 90747}
\newcommand*{\CSUDHindex}{3}
\affiliation{\CSUDH}
\newcommand*{\CMU}{Carnegie Mellon University, Pittsburgh, Pennsylvania 15213}
\newcommand*{\CMUindex}{4}
\affiliation{\CMU}
\newcommand*{\CUA}{Catholic University of America, Washington, D.C. 20064}
\newcommand*{\CUAindex}{5}
\affiliation{\CUA}
\newcommand*{\SACLAY}{Irfu/SPhN, CEA, Universit\'e Paris-Saclay, 91191 Gif-sur-Yvette, France}
\newcommand*{\SACLAYindex}{6}
\affiliation{\SACLAY}
\newcommand*{\CNU}{Christopher Newport University, Newport News, Virginia 23606}
\newcommand*{\CNUindex}{7}
\affiliation{\CNU}
\newcommand*{\UCONN}{University of Connecticut, Storrs, Connecticut 06269}
\newcommand*{\UCONNindex}{8}
\affiliation{\UCONN}
\newcommand*{\FU}{Fairfield University, Fairfield CT 06824}
\newcommand*{\FUindex}{9}
\affiliation{\FU}
\newcommand*{\FERRARAU}{Universita' di Ferrara , 44121 Ferrara, Italy}
\newcommand*{\FERRARAUindex}{10}
\affiliation{\FERRARAU}
\newcommand*{\FIU}{Florida International University, Miami, Florida 33199}
\newcommand*{\FIUindex}{11}
\affiliation{\FIU}
\newcommand*{\FSU}{Florida State University, Tallahassee, Florida 32306}
\newcommand*{\FSUindex}{12}
\affiliation{\FSU}
\newcommand*{\GWUI}{The George Washington University, Washington, DC 20052}
\newcommand*{\GWUIindex}{13}
\affiliation{\GWUI}
\newcommand*{\ISU}{Idaho State University, Pocatello, Idaho 83209}
\newcommand*{\ISUindex}{14}
\affiliation{\ISU}
\newcommand*{\INFNFE}{INFN, Sezione di Ferrara, 44100 Ferrara, Italy}
\newcommand*{\INFNFEindex}{15}
\affiliation{\INFNFE}
\newcommand*{\INFNFR}{INFN, Laboratori Nazionali di Frascati, 00044 Frascati, Italy}
\newcommand*{\INFNFRindex}{16}
\affiliation{\INFNFR}
\newcommand*{\INFNGE}{INFN, Sezione di Genova, 16146 Genova, Italy}
\newcommand*{\INFNGEindex}{17}
\affiliation{\INFNGE}
\newcommand*{\INFNRO}{INFN, Sezione di Roma Tor Vergata, 00133 Rome, Italy}
\newcommand*{\INFNROindex}{18}
\affiliation{\INFNRO}
\newcommand*{\INFNTUR}{INFN, Sezione di Torino, 10125 Torino, Italy}
\newcommand*{\INFNTURindex}{19}
\affiliation{\INFNTUR}
\newcommand*{\ORSAY}{Institut de Physique Nucl\'eaire, CNRS/IN2P3 and Universit\'e Paris Sud, Orsay, France}
\newcommand*{\ORSAYindex}{20}
\affiliation{\ORSAY}
\newcommand*{\ITEP}{Institute of Theoretical and Experimental Physics, Moscow, 117259, Russia}
\newcommand*{\ITEPindex}{21}
\affiliation{\ITEP}
\newcommand*{\JMU}{James Madison University, Harrisonburg, Virginia 22807}
\newcommand*{\JMUindex}{22}
\affiliation{\JMU}
\newcommand*{\KNU}{Kyungpook National University, Daegu 41566, Republic of Korea}
\newcommand*{\KNUindex}{23}
\affiliation{\KNU}
\newcommand*{\LPSC}{LPSC, Universit\'e Grenoble-Alpes, CNRS/IN2P3, Grenoble, France}
\newcommand*{\LPSCindex}{24}
\affiliation{\LPSC}
\newcommand*{\MISS}{Mississippi State University, Mississippi State, MS 39762-5167}
\newcommand*{\MISSindex}{25}
\affiliation{\MISS}
\newcommand*{\UNH}{University of New Hampshire, Durham, New Hampshire 03824-3568}
\newcommand*{\UNHindex}{26}
\affiliation{\UNH}
\newcommand*{\NSU}{Norfolk State University, Norfolk, Virginia 23504}
\newcommand*{\NSUindex}{27}
\affiliation{\NSU}
\newcommand*{\OHIOU}{Ohio University, Athens, Ohio  45701}
\newcommand*{\OHIOUindex}{28}
\affiliation{\OHIOU}
\newcommand*{\ODU}{Old Dominion University, Norfolk, Virginia 23529}
\newcommand*{\ODUindex}{29}
\affiliation{\ODU}
\newcommand*{\RPI}{Rensselaer Polytechnic Institute, Troy, New York 12180-3590}
\newcommand*{\RPIindex}{30}
\affiliation{\RPI}
\newcommand*{\URICH}{University of Richmond, Richmond, Virginia 23173}
\newcommand*{\URICHindex}{31}
\affiliation{\URICH}
\newcommand*{\ROMAII}{Universita' di Roma Tor Vergata, 00133 Rome Italy}
\newcommand*{\ROMAIIindex}{32}
\affiliation{\ROMAII}
\newcommand*{\MSU}{Skobeltsyn Institute of Nuclear Physics, Lomonosov Moscow State University, 119234 Moscow, Russia}
\newcommand*{\MSUindex}{33}
\affiliation{\MSU}
\newcommand*{\SCAROLINA}{University of South Carolina, Columbia, South Carolina 29208}
\newcommand*{\SCAROLINAindex}{34}
\affiliation{\SCAROLINA}
\newcommand*{\TEMPLE}{Temple University,  Philadelphia, PA 19122 }
\newcommand*{\TEMPLEindex}{35}
\affiliation{\TEMPLE}
\newcommand*{\JLAB}{Thomas Jefferson National Accelerator Facility, Newport News, Virginia 23606}
\newcommand*{\JLABindex}{36}
\affiliation{\JLAB}
\newcommand*{\UTFSM}{Universidad T\'{e}cnica Federico Santa Mar\'{i}a, Casilla 110-V Valpara\'{i}so, Chile}
\newcommand*{\UTFSMindex}{37}
\affiliation{\UTFSM}
\newcommand*{\EDINBURGH}{Edinburgh University, Edinburgh EH9 3JZ, United Kingdom}
\newcommand*{\EDINBURGHindex}{38}
\affiliation{\EDINBURGH}
\newcommand*{\GLASGOW}{University of Glasgow, Glasgow G12 8QQ, United Kingdom}
\newcommand*{\GLASGOWindex}{39}
\affiliation{\GLASGOW}
\newcommand*{\UKL}{University of Kentucky, Lexington, KY 40506}
\newcommand*{\UKLindex}{40}
\affiliation{\UKL}
\newcommand*{\VCU}{Virginia Commonwealth University, Richmond, Virginia 23284}
\newcommand*{\VCUindex}{41}
\affiliation{\VCU}
\newcommand*{\VT}{Virginia Tech, Blacksburg, Virginia   24061-0435}
\newcommand*{\VTindex}{42}
\affiliation{\VT}
\newcommand*{\VIRGINIA}{University of Virginia, Charlottesville, Virginia 22901}
\newcommand*{\VIRGINIAindex}{43}
\affiliation{\VIRGINIA}
\newcommand*{\WM}{College of William and Mary, Williamsburg, Virginia 23187-8795}
\newcommand*{\WMindex}{44}
\affiliation{\WM}
\newcommand*{\YEREVAN}{Yerevan Physics Institute, 375036 Yerevan, Armenia}
\newcommand*{\YEREVANindex}{45}
\affiliation{\YEREVAN}
\newcommand*{\HAMPTON}{Hampton University, Hampton, VA 23668}
\newcommand*{\HAMPTONindex}{46}
\affiliation{\HAMPTON}
\newcommand*{\LANL}{Los Alamos National Laboratory, Los Alamos, NM 87544 USA}
\newcommand*{\LANLindex}{47}
\affiliation{\LANL}

\author{P.E.~Bosted}
     \email{bosted@jlab.org}
\affiliation{\WM}
\author {A.~Kim} 
\affiliation{\UCONN}
\author {K.P. ~Adhikari} 
\affiliation{\MISS}
\affiliation{\ODU}
\author {D.~Adikaram} 
\affiliation{\JLAB}
\affiliation{\ODU}
\author {Z.~Akbar} 
\affiliation{\FSU}
\author {M.J.~Amaryan} 
\affiliation{\ODU}
\author {S. ~Anefalos~Pereira} 
\affiliation{\INFNFR}
\author {H.~Avakian} 
\affiliation{\JLAB}
\author {R.A.~Badui} 
\affiliation{\FIU}
\author {J.~Ball} 
\affiliation{\SACLAY}
% ...

\collaboration{The CLAS Collaboration}
\noaffiliation

\date{\today}


\begin{abstract}
Beam-target double-spin asymmetries and target single-spin
asymmetries were measured for the 
exclusive  $\pi^0$ electroproduction reaction
$\gamma^* p \to p \pi^0$, expanding an analysis of
the $\gamma^* p \to n \pi^+$ reaction from the same experiment.
The results were obtained from scattering of 6~GeV 
longitudinally polarized
electrons off longitudinally polarized protons
using the CEBAF Large Acceptance Spectrometer 
at Jefferson Lab. 
The kinematic range covered is $1.1<W<3$ GeV and $1<Q^2<6$
GeV$^2$. Results were obtained for about 5700
bins in $W$, $Q^2$, \cthcm, and $\phi^*$. The beam-target
asymmetries were found to generally be greater than zero,
with relatively modest \phicmsp dependence. The target
asymmetries exhibit very strong \phicmsp dependence,
with a change in sign occurring between results at
low $W$ and high $W$, in contrast to $\pi^+$ electroproduction.
 Reasonable agreement is
found with phenomenological fits to previous data
for $W<1.6$ GeV, but significant  differences are seen
at higher $W$.  When combined with 
cross section measurements, as well as $\pi^+$ observables,
the present results will provide powerful constraints
on nucleon resonance amplitudes at moderate and large
values of $Q^2$, for resonances with masses as high as 2.4 GeV.   
\end{abstract}
\maketitle


\section{Introduction}
This article is a companion to a previous 
publication~\cite{eg1cpip}, which presents data for the
target and beam-target spin asymmetries in exclusive 
$\pi^+$ electroproduction for $Q^2>1$ GeV$^2$. The present
article expands upon~\cite{eg1cpip} to provide results
for $\pi^0$ electroproduction. Briefly, the physics
motivation is to study nucleon structure and reaction
mechanisms via large-$Q^2$ pion electroproduction. 
The results are from the ``eg1-dvcs'' experiment, which
used scattering of 6~GeV longitudinally polarized
electrons off longitudinally polarized protons. Scattered
electrons and electroproduced neutral pions were detected
in the CEBAF Large Acceptance Spectrometer~\cite{CLAS} (CLAS) 
at Jefferson Lab, augmented for this experiment with
an Inner Calorimeter (IC). This
calorimeter consists of an array of small lead-tungstate
crystals, each 15 cm long and roughly 2 cm square. 
The IC greatly increased the acceptance for neutral pions
compared to the standard setup.  The primary target for this
analysis consisted of a 1.5-cm-long cell with about 1 g/cm$^2$
of ammonia immersed in a liquid-helium bath. An auxiliary
target with carbon instead of ammonia was used for background
studies. The data taking relevant to the present analysis 
was divided into two parts, for which the target position,
electron beam energy, and beam and target polarizations
are listed in Table~\ref{tab:parts}.

\begin{table}[hbt]
\begin{tabular}{lcccc}
Run Period & $z$ & \> $E$ & \> $P_BP_T$ & \> $P_B$ \\
\hline
Part A     & -58 cm & \> 5.887 GeV   & \> $0.637\pm 0.011$ & \> $0.85\pm0.04$ \\
Part B     & -68 cm & \> 5.954 GeV   & \> $0.645\pm 0.007$ & \> $0.85\pm0.04$\\
\hline
\end{tabular}
\caption{Run period names, target position along the beam line
relative to CLAS center ($z$), nominal beam energy ($E$), 
$P_BP_T$ and $P_B$, where $P_B$ ($P_T$) is
the beam (target) polarization, for the two running periods of
the experiment.
}
\label{tab:parts}
\end{table}

 For further elucidation of
the physics motivation, details on the formalism, 
experimental overview, and details on the detection of
scattered electrons, please see the companion article~\cite{eg1cpip}
as well as other publications from the eg1-dvcs 
experiment on inclusive electron scattering~\cite{inclprc}
and Deep Virtual Compton Scattering \cite{dvcsprd}.

Large four-momentum transferred $Q^2$ measurements 
of spin-averaged 
cross sections for exclusive $\pi^0$ electroproduction 
from a proton are sparse compared to $\pi^+$ production, and published
results are limited to the $\Delta(1232)$ resonance
region~\cite{frolov,villano}, with results at higher 
invariant mass $W$ from
CLAS still under analysis~\cite{Ungaro}, although the 
beam single-spin asymmetries ($A_{LU}$) were 
published \cite{Masi} several years ago. 
Beam-target asymmetries ($A_{LL}$) and target single-spin 
asymmetries ($A_{UL}$)
for \tpIIr were reported from the ``eg1b" experiment
at Jefferson Lab~\cite{Biselli} at relatively low $Q^2$
for an electron beam energy of 1.7 GeV. Results for
$A_{LL}$ and $A_{UL}$ at much larger values of $Q^2$ from
the present experiment were reported in Ref.~\cite{kim}, 
for values of the final state invariant mass $W$ above 2 GeV.
The present analysis expands upon Ref.~\cite{kim} 
to include $W<2$ GeV and provide higher statistical
precision for $W>2$ GeV through the inclusion of additional
final state topologies.

\section{Analysis}

The data analysis for $\pi^0$ electroproduction proceeded
in parallel with that for $\pi^+$ 
electroproduction as described in the companion article 
Ref.~\cite{eg1cpip}.
 
\subsection{Particle identification}
We analyzed 
$\pi^0$ electroproduction using three topologies:
\tpIIn, \tpVI and \tpIX. No event was counted in more than
one topology. All three topologies require detection of
the scattered electron and at least one photon. The \tpII
and \tpIX topologies require the detection of two photons with
invariant mass corresponding to a $\pi^0$. The \tpII and \tpVI
topologies also require the detection of a proton.  The
cuts used to identify scattered electrons are given in 
Ref.~\cite{eg1cpip}.

\subsubsection{Proton identification}
Protons were identified
by requiring a positively charged track with a time-of-arrival
at the scintillation counters within 0.7 ns (approximately 3$\sigma$) 
of that predicted from
the time-of-arrival of the electron in the event.
This timing cut removed all charged pions from the sample,
but allowed between 10\% to 100\% of $K^+$, depending
on kaon momentum. These events were removed by the
missing mass cut discussed below. Positrons were
removed from the sample by requiring small (or no)
signal in the Cherenkov detector and a 
small deposited energy in the electromagnetic calorimeter (EC).
Also required were a vertex position
reconstructed (with a resolution of 5 to 8 mm) 
within 4 cm of the nominal target center,
and a polar scattering angle between 15 and 48 degrees.

\subsubsection{Photon identification}
Photons in the EC were identified with the 
following criteria:
no associated track (to ensure neutrality); 
energy greater than 0.3 GeV (to have sufficiently good
energy resolution); 
time-of-arrival at the EC in agreement with the scattered
electron time within 3 ns (to reduce the rate
of accidental coincidences); and an anti-Bremsstrahlung
cut of 3.4 degrees. A photon was considered to be
a candidate for Bremsstrahlung from the
scattered electron if the opening angle between
the electron and the photon was less than 3.4 degrees
at either the target vertex, or at the first
drift chamber. The reason that both places were checked
is that the electron undergoes a significant
azimuthal rotation in the target solenoid. 

Photons in the IC were identified
by requiring a deposited energy of at least
0.2 GeV (to ensure adequate
energy resolution) and a time-of-arrival within 2 ns of that
calculated from the scattered electron arrival time
(to reject random background). 
Single photons in the IC (for the
topology \tpVIn) were not considered, because exclusivity
cuts in this case were not sufficient to ensure that the
missing particle was really a photon from $\pi^0$ decay.
It was found that there was a large background of events
in which the IC particle was an electron (rather than a photon), 
and the missing particle was a positron, i.e. 
$$e^- p \rightarrow e^- \> {\rm (in \> IC)} \> p \> {\rm (in \> CLAS)} \> 
 e^- \> {\rm (in \> CLAS)} \> (e^+, \> {\rm missing}).$$
In this case, the electron in CLAS and the missing positron are
the products of the decays of  $\pi^0$, $\eta$, or other mesons.

\subsubsection{$\pi^0$ identification}
\begin{figure}[hbt]
\centerline{\includegraphics[width=3.2in,angle=90]{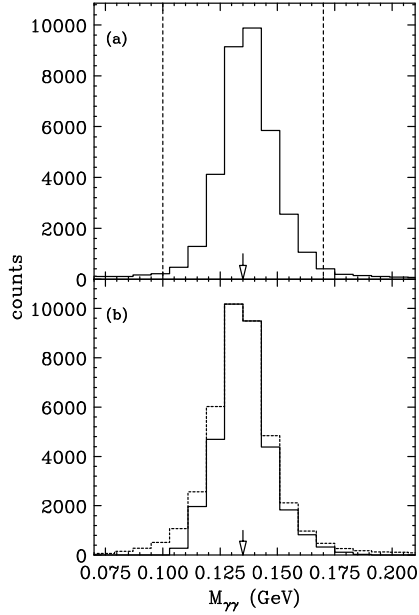}}
\caption{Two-photon invariant mass distributions for 
a) \tpII and b) \tpIXn, with all
relevant exclusivity cuts applied. 
The vertical dashed lines show the cuts used for \tpIIn.
 The solid (dashed) curve in the lower panel is 
with (without) the application of the $\chi^2$
cut discussed in the text.
}
\label{fig:m2g}
\end{figure}

For topologies \tpII and \tpIXn, a $\pi^0$ was identified 
using the invariant mass of the photon pair. 
Fig.~\ref{fig:m2g}a shows the mass 
distributions for events passing all other exclusivity
cuts for the topology \tpIIn.
The background under the peak is very small
(less than 1.5\%) for this topology.
The vertical dashed lines show the cuts used:
$0.10<M_{\gamma\gamma}<0.17 \> {\rm GeV}$.

The two-photon mass distribution for  topology \tpIX
is shown in Fig.~\ref{fig:m2g}b.
The dashed curve is for events passing the 
electron-meson missing mass cut discussed below. 
There is more background under the $\pi^0$ peak
than for topology \tpII (as evidenced by the enhancement around
0.1 GeV). Rather than using a simple
two-photon mass cut, it was found that a more complicated
cut was better at removing background events. 
 The solid curve is with
the cut $\chi^2<4$, where $\chi^2$ is defined in the next
paragraph. The cut value was chosen to optimize the signal-to-noise.

In order to get the best possible determination of 
electron-pion missing mass, we adjusted the energy of
each of the two photons such that the invariant mass
was exactly equal to the $\pi^0$ mass $M_0$. We did
not adjust the photon angles, because the energy resolution
is the dominant contribution. We can define
\be
M_0^2 / M_{\gamma\gamma}^2 = (1 + c_1\sigma_1)(1+c_2\sigma_2),
\ee
where $M_{\gamma\gamma}$ is the measured invariant two-photon
mass, $c_1$ and $c_2$ are coefficients to be determined
by minimizing $\chi^2=c_1^2+c_2^2$, and the relative photon
energy resolutions $\sigma_i$ were approximated by
$$\sigma_i = 0.01 + \frac{0.05}{\sqrt{E_\gamma}} 
\> {\rm for \> IC}$$
$$\sigma_i = 0.02 + \frac{0.12}{\sqrt{E_\gamma}} 
\> {\rm for \> EC}.$$
After the fit was done, the photon energies were scaled
by $(1+c_i\sigma_i)$.
 
\subsection{Exclusivity kinematic cuts}

For all three topologies, kinematic cuts
were placed to improve the signal to noise ratio. The value
of kinematic cuts is two-fold. First,
most of the kinematic quantities have a wider
distribution for bound nucleons (in target materials
with atomic number $A>2$) than for free protons. Kinematic cuts therefore
reduce the dilution of the signal of interest
(scattering from polarized free protons) compared to the
background from unpolarized nucleons in materials
with $A>2$. Second, kinematic cuts are needed to
isolate single meson production from multi-meson
production. Multi-meson production was further reduced
by eliminating events in which any extra particles
were detected in CLAS or the IC.


\subsubsection{Electron-pion missing mass cut}
For both the \tpII and \tpIX topologies, 
the electron-pion missing mass \mxepi
should be equal to the proton mass of 0.938 GeV.
In general, one would like the upper cut on \mxepi
to be well below  $M+m_\pi=1.08$ GeV, to avoid
contributions from multi-pion production. Placing
tighter cuts helps to reduce the nuclear background.

\begin{figure}[hbt]
\centerline{\includegraphics[width=3.2in,angle=90]{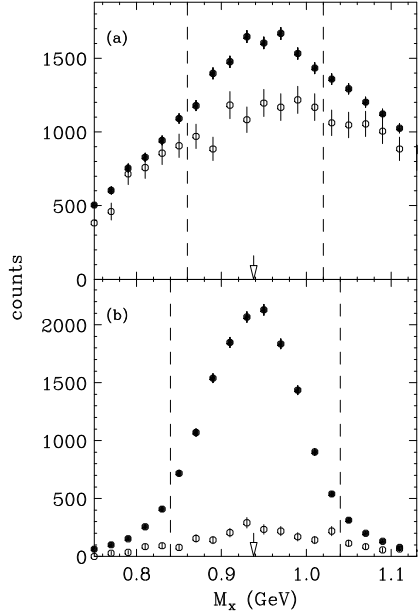}}
\caption{Electron-pion missing mass spectra for the
topologies a) \tpIX and b) \tpIIn. 
Counts from the ammonia target are shown
as the solid circles and counts from the carbon target
(scaled by the ratio of integrated luminosities on 
bound nucleons) are shown as the open circles.  
All other applicable exclusivity cuts have been applied. 
The vertical dashed lines indicate the cuts used.
}
\label{fig:www}
\end{figure}

The distribution in  \mxepi is shown for
the fully exclusive topology 
\tpII in Fig.~\ref{fig:www}b
averaged over the full kinematic range of the experiment. 
All other applicable exclusivity cuts have been applied.
The solid circles correspond to counts from the ammonia
target, while the open circles correspond to counts
from the carbon target, scaled by the ratio of
luminosities on $A>2$ nucleons. A clear
peak is seen near the nucleon mass from the ammonia
target, with a smaller but wider distribution from the
carbon target, that matches the wings on the ammonia
distributions on the low-mass side of the peak. On
the high side of the peak, the ammonia rates are
higher, due to the radiative tail of the single-pion
production, and the gradual turn-on of multi-pion
production. The vertical dashed lines show the
cuts used: $0.86<M_x^{e\pi}<1.04$ GeV. Within the
cut region, approximately 10\% of the events come
from nucleons in nuclei with $A > 2$ and 90\% from free protons. 

The distribution in  \mxepi is shown for topology
\tpIX in Fig.~\ref{fig:www}b,
for $W<1.5$ GeV. 
The nuclear background is considerably larger
in this case, because there are no other exclusivity
cuts that can be applied for this topology. For this
reason, we used tighter missing mass cuts of 
$0.88<M_x^{e\pi}<1.02$ GeV. For $W>1.5$ GeV, an
increasingly large multi-pion background was observed,
and those events were not used in the analysis.  

The spectra were examined to
see if the optimal cut values depends on $W$,
$Q^2$, \cthcm, or \phicm. Although the peak widths
vary somewhat with kinematic variables, a constant cut value did not
degrade the signal to noise ratios by more than a few percent. 

\subsubsection{Electron-proton missing mass cuts}
\begin{figure}[hbt]
\centerline{\includegraphics[width=3.2in,angle=90]{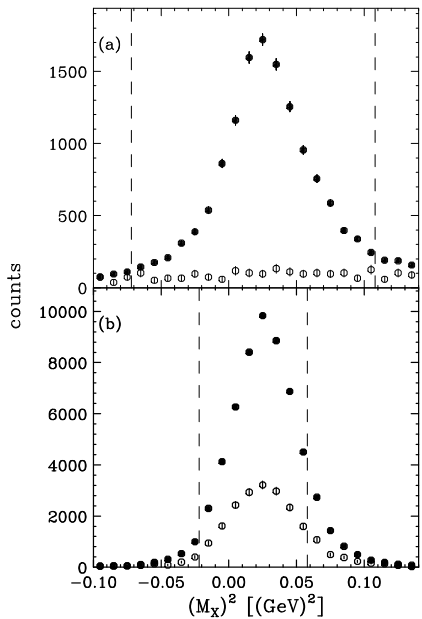}}
\caption{Distribution of $(M_x^{eN})^2$ for: a) the
topology \tpIIn; and b) the topology \tpVIn. 
Symbols are as in Fig.~\ref{fig:www}.
The vertical dashed lines show the cuts used.
All other relevant exclusivity cuts have been applied.
}
\label{fig:wwwep}
\end{figure}

In the two topologies for which a proton was measured in the
final state, the squared electron-proton missing mass $(M_x^{eN})^2$
should equal the $\pi^0$ mass squared. The
spectra for the two topologies are shown
in Fig.~\ref{fig:wwwep}, averaged over the kinematic
range of the experiment. The cuts are shown as the
vertical dashed lines, and correspond to
$ 0.07<(M_x^{eN})^2<0.11$ GeV$^2$ for
topology \tpII and 
$ 0.02<(M_x^{eN})^2<0.06$ GeV$^2$ for topology \tpVIn.
These cuts are very effective in reducing
nuclear background, as well as eliminating multi-meson
production. The larger tails at positive values of $(M_x^{eN})^2$
are the result of photon radiation by the incoming or scattered electron.

\subsubsection{Proton angular cuts}
\begin{figure}[hbt]
\centerline{\includegraphics[width=3.2in,angle=90]{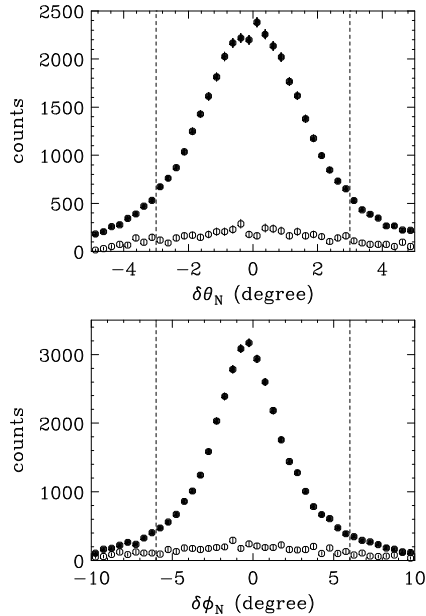}}
\caption{Distribution of the in-plane (out-of-plane) 
angular difference in the
predicted and observed proton direction cosines 
for the topology \tpII are shown in the upper (lower) panel.
The solid black points are for the ammonia target, while
the open circles are from the carbon target, scaled
by integrated luminosity. 
The vertical dashed lines
indicate the cuts used in the analysis. 
All other relevant exclusivity cuts have been applied.
}
\label{fig:dthphi}
\end{figure}

In the topology \tpIIn, cuts on the cone angles of 
the detected proton
are useful in 
rejecting background from $A>2$ materials. From the
kinematics of the detected electron and pion, the
direction cosines of the recoil proton are calculated,
and compared with the observed angles. We denote
the difference in predicted and observed angles
as $\delta \theta_N$ in the in-plane direction and
$\delta \phi_N$ in the out-of-plane direction (which
tends to have worse experimental resolution). Distributions
of these two quantities are shown
in Fig.~\ref{fig:dthphi}. It can be seen that with 
cuts on $M_x$ and
the complementary angle, the nuclear background is relatively small 
and flat compared to the peaks from the free proton.  
We used the cuts $| \delta \theta_N | <3^\circ$
and $| \delta \phi_N | <6^\circ$, for all kinematic bins. 

\subsubsection{Specific cuts used for topology \tpVIn}
\begin{figure}[hbt]
\centerline{\includegraphics[width=3.2in,angle=90]{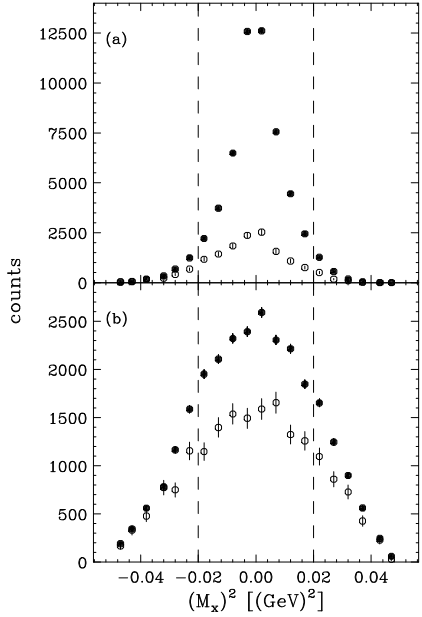}}
\caption{Distributions of electron-proton-photon
missing mass squared for the \tpVI topology
for a) $1.1<W<1.45$ GeV and 
b) $2.15<W<2.5$ GeV. 
 Symbols are as in Fig.~\ref{fig:www}. The vertical dashed
lines show the cuts used. 
All other relevant exclusivity cuts have been applied.
}
\label{fig:wwwp}
\end{figure}

Four cuts were applied for the \tpVI
topology. The first was to require that the 
electron-proton-photon missing mass squared
$(\mxepg)^2$ be close to zero, to ensure that the missing
particle (if any), is a photon. The spectra
at low and high $W$ values are shown 
in Fig.~\ref{fig:wwwp}, along with the cut:
$-0.02< (\mxepg)^2  <0.02 \> {\rm GeV}^2$.

Two cuts for \tpVI were used to reduce contamination
from events from the virtual Compton
Scattering (VCS) reaction $e p \rightarrow e p \gamma$.
The VCS reaction differs from $\pi^0$ production
by: a) electron-proton-photon missing energy $E_{miss}$=0;
b) the difference
in angle between the observed photon, and the angle 
predicted from the detected electron and proton
$\delta\theta_\gamma=0$, while
for $\pi^0$ production, both of 
these quantities are positive. In addition,
the photon energy on average is much
larger for VCS than for $\pi^0$ production.

\begin{figure}[hbt]
\centerline{\includegraphics[width=3.2in,angle=90]{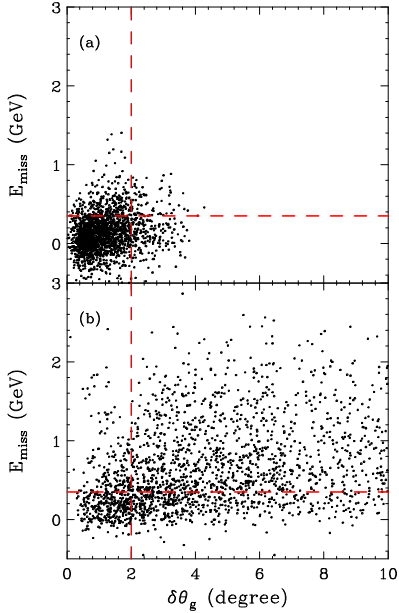}}
\caption{Distributions of angular difference between the
predicted and measured photon (horizontal axis) versus
electron-proton-photon missing energy (vertical axis)
for the the \tpVI topology.
Panel a) is for photons with energy greater 
than 2 GeV, with the remainder of the events in panel b).
The dashed lines indicate the
two cuts used in the analysis. All other exclusivity
cuts have been applied. 
}
\label{fig:dedthg}
\end{figure}

The features of VCS events can be readily
seen in Fig.~\ref{fig:dedthg} as a strong enhancement 
at small values of both $\delta\theta_g$ and $E_{miss}$,
especially for events with photon energies 
 greater than 2 GeV (Fig.~\ref{fig:dedthg}a), 
with weaker population in this region for lower photon
energies (Fig.~\ref{fig:dedthg}b). 
The dashed lines indicate the cuts used in the analysis.
The cuts were applied  differently for high and low
photon energies:
\be
\delta\theta_g>2 \> {\rm degrees} \> {\rm AND} \> 
E_{miss}>0.35 \> {\rm GeV \> for} \, 
E_\gamma > 2 \> {\rm GeV}
\ee
\be
\delta\theta_g>2 \> {\rm degrees} \> {\rm OR} \> 
E_{miss}>0.35 \> {\rm GeV \> for} \, 
E_\gamma < 2 \> {\rm GeV}.
\ee

Another cut was used to reject events where the actual reaction
was not from electron scattering, but rather a photoproduction reaction, i.e.
$\gamma p \rightarrow p e^- \gamma (e^+)$, where the
$\gamma$, $e^-$, and missing $e^+$ come from $\pi^0$ Dalitz decay. In this
case, the opening angle between the electron and positron is
zero. Such events result in an enhancement in the
difference in azimuthal angles between the measured electron
and the missing positron (calculated assuming the missing particle
is a photon). We put a cut
of $\pm 30$ degrees to eliminate these rarely-occurring events.

\begin{figure}[hbt]
\centerline{\includegraphics[width=3.2in,angle=90]{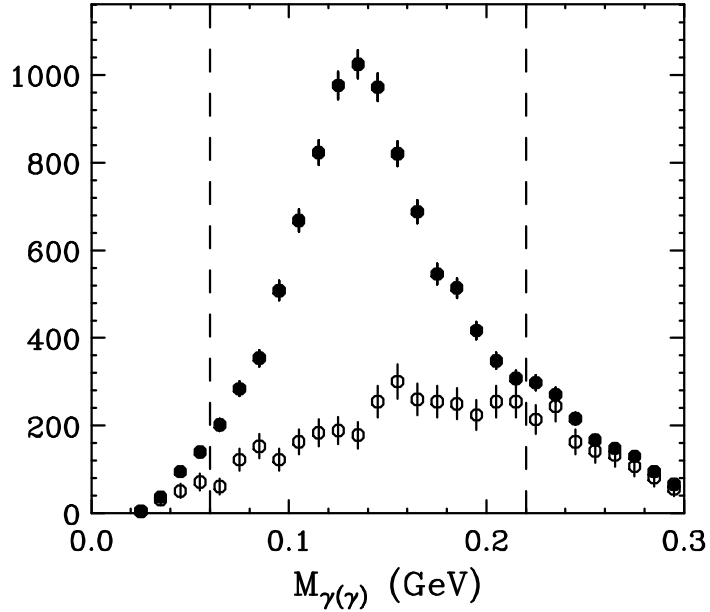}}
\caption{Distribution of $M_{\gamma (\gamma)}$ for
the topology \tpVIn.  
 Symbols are as in Fig.~\ref{fig:www}. The vertical dashed
lines show the cuts used. 
All other relevant exclusivity cuts have been applied.
}
\label{fig:m2g6}
\end{figure}

The final cut was on the quantity 
$M_{\gamma (\gamma)}$, which is the invariant mass of 
the detected photon and the missing particle, with
the imposed constraint that the mass of the missing
particle is zero. As shown in Fig.~\ref{fig:m2g6}, 
the $M_{\gamma (\gamma)}$ spectrum is consistent
with pure neutral pion production when all other 
exclusivity cuts are applied. We used the cut
$0.06 < M_{\gamma (\gamma)} <0.22$ GeV.

\subsubsection{Additional cuts}
For topology \tpIIn,
the energy of all final state particles is measured,
and therefore the missing energy $E_m$ distribution is
centered on zero for free proton events, and about 0.02
GeV for bound protons. A cut of $E_m<0.13$ GeV was used
to give a slight improvement in the signal-to-noise ratio.
For topology \tpIXn, only events with $W<1.5$~GeV
were used, as mentioned above. For topology \tpVIn,
only events with the photon in the EC were used. 
 
\subsection{Kinematic binning}\label{sec:kin}

The kinematic
range of the experiment is $1.1<W<3$ GeV and 
$1<Q^2<6$ GeV$^2$. As shown in Fig.~1 of
Ref.~\cite{eg1cpip}, the range in $Q^2$ changes with $W$. 
We therefore
made four bins in $Q^2$, where the
limits correspond to electron scattering angles
of 15.5, 18, 21, 26, and 38 degrees. We 
used fixed $W$ bins of width 0.05 GeV for $W<1.9$ GeV,
which is comparable to the experimental 
resolution. For $W>1.9$ GeV, the bin widths gradually
increase to achieve roughly equal counting rates, with
bin boundaries at 1.90, 1.96, 2.03, 2.11, 2.20, 2.31, 2.43, 2.56,
2.70, 2.85 and 3 GeV.        
We used six bins in $\cos(\theta^*)$, with
boundaries at -0.6, -0.2, 0.1, 0.36, 0.6, 0.85, and 0.995.
We used 12 bins in $\phi^*$, equally spaced
between 0 and $2\pi$. 

A strong consideration in choosing the bin sizes was that
we required at least ten counts in a given bin in order
to have approximately Gaussian statistical uncertainties. 
The total number of bins is 7488, of which about 5700
had enough events to be included in the final results.

\newpage\section{Asymmetries}
Spin asymmetries were formed as follows:
\be
A_{LL} = \frac{
   N^{\uparrow\downarrow} + 
   N^{\downarrow\uparrow} -
   N^{\uparrow\uparrow} - 
   N^{\downarrow\downarrow}}{
   N_{tot} \hskip .05in f \hskip .05in P_BP_T},
\ee
\be
A_{UL} = \frac{
   N^{\uparrow\uparrow} + 
   N^{\downarrow\uparrow} -
   N^{\uparrow\downarrow} - 
   N^{\downarrow\downarrow}}{
   N_{tot} \hskip .05in f \hskip .05in P_T},
\ee
where the symbols $N$ represent the number of events
in a given helicity configuration, divided by the
corresponding integrated beam current. The first superscript
refers to the beam polarization direction, and the
second to the target polarization direction. The total
number of counts is denoted by
$N_{tot}=N^{\uparrow\uparrow} + 
               N^{\downarrow\uparrow} +
               N^{\uparrow\downarrow} + 
               N^{\downarrow\downarrow}$,
and $f$ is the dilution factor, defined as the fraction
of events originating from polarized free protons, compared
to the total number of events.
The product of beam polarization ($P_B$) and
target polarization ($P_T$), 
as well as the value of $P_B$, are listed
in Table I for the two Parts of the experiment.

\subsection{Dilution factor}
\label{sec:df}
The dilution factor $f$ is defined as the ratio of 
scattering rate from free nucleons to the scattering 
rate from all nucleons in the target.
With the assumption that the cross section per nucleon
is the same for bound protons in all of the nuclear
materials (with $A>2$) in a given target, and also that
the effective detection efficiency is the same for the
ammonia and carbon targets, then
\be
f = 1 - R_{A>2} \frac{N_C}{N_{NH_3}},
\label{Eq:f}
\ee
where 
$N_C$ and $N_{NH_3}$ are the number of counts from the
carbon and ammonia targets respectively,
measured in a given kinematic bin for a given topology,
normalized by the corresponding integrated beam charge. The
symbol $R_{A>2}$ denotes the ratio of the number of bound nucleons
in the  ammonia target to the number of bound nucleons in the
carbon target. Bound nucleons are defined to be in
materials with atomic number $A>2$.
The latter was determined from a detailed analysis of the
target composition using inclusive electron scattering
rates from ammonia, carbon, and 
empty targets, yielding
$R_{A>2}=0.71$ for Part A and $R_{A>2}=0.72$ for Part B.

Because the integrated luminosity on the carbon target
was about ten times lower than on the ammonia
target, there is a large amplification of the uncertainty
on the ratio of carbon to ammonia counts, 
$\frac{N_C}{N_{NH_3}}$. In many cases, this would lead
to unphysical values of $f$ (i.e. $f<0$). We therefore
took advantage of the fact that $f$ is a very slowly
varying function of kinematic variables, and did a global fit
to $\frac{N_C}{N_{NH_3}}$. 
The fit values were then
used to evaluate $f$ in each kinematic bin.

As in Ref.~\cite{eg1cpip}, the functional forms for the fit 
contained 25 terms of the
form $p_i \cos^{N_c}(\theta^*) W^{N_W} (Q^2)^{N_Q}$, where 
$p_i$ is a free parameter, and the exponents $N_C$, $N_W$, and
$N_Q$ range from 0 to 3 (although not all possible terms were
included). An additional eight terms were included to 
account for the influence
of the three prominent nucleon resonances centered at 
$1.23$ GeV, 
$1.53$ GeV, and 
$1.69$ GeV,
all with widths of $0.120$ GeV.
The reason that these resonance terms are
needed is that the nucleon resonances are effectively broadened
in the target materials with $A>2$ by Fermi motion.
This generates resonant-like structures in
the ratio of carbon to ammonia count rates. 
Tests were made to see if any $\phi^*$-dependent
terms would improve the fits. No significant 
improvements were found. 

\begin{figure}[hbt]
\centerline{\includegraphics[width=3.2in,angle=90]{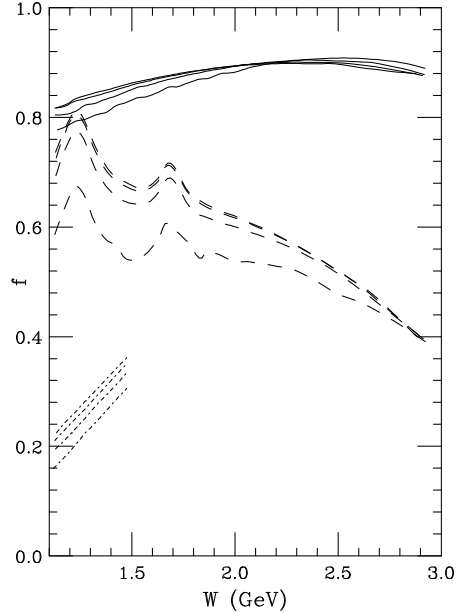}}
\caption{Dilution factors as a function of $W$ for
the \tpII topology (solid curves), the \tpVI topology
(long dashed curves), and the \tpIX topology
(short dashed curves) for the four $Q^2$ bins of 
this experiment and a typical bin in $\cos(\theta^*)$.
For the two sets of dashed curves, smaller values of $f$
correspond to higher values of $Q^2$.
}
\label{fig:dilff1}
\end{figure}

The dilution factors for Part B for 
the three topologies are shown in 
Fig.~\ref{fig:dilff1} as a function of $W$ for the four $Q^2$
bins of this analysis and a typical bin in \cthcm. 
 For the fully
exclusive topology, \tpIIn, the dilution factor
is large, about 0.85 on average, corresponding
to the good rejection of background that is possible
with the exclusivity cuts when the recoil proton
is detected.  For the topology \tpVIn, the dilution 
factor is reasonably good for $W<2$ GeV,
averaging about 0.65, with significant resonant 
structure visible. For $W>2$ GeV, there is a
trend for $f$ to decrease, dropping to values as 
low as 0.4 at the highest values of $W$. This is
because Fermi broadening results in an increasing amount
 of multi-pion production from
the nuclear target material. The dilution factor
for topology \tpIX is much lower than for the other
two topologies, averaging about 0.25.
The $Q^2$-dependence is 
relatively weak, although there is a trend towards lower
values of $f$ at higher values of $Q^2$. 
Because Part A had much lower statistical accuracy than
Part B, we used the Part B fits for Part A.

\subsection{Combining data sets}
The entire asymmetry analysis was performed separately
for Part A and Part B.
The results were combined by 
averaging asymmetries, weighted by their 
respective statistical uncertainties, for each of the
4-dimensional bins. Since the two configurations
differ only in the acceptance function, which should
cancel in forming the asymmetries, the
expectation is that the acceptance functions should 
be fully compatible statistically.
This expectation was verified for 
both asymmetries for all three topologies.

\subsection{Combining topologies}

We next averaged together the asymmetry results for the
three topologies, weighted by their 
respective statistical uncertainties, for each of the
4-dimensional bins.
For both  asymmetries, the
topologies were found to be statistically compatible,
indicating that the dilution factors
for the different topologies are properly accounted for.
We found that topology \tpII is the biggest contributor
at high $W$, while topology \tpVI dominates at low
values of $W$. Due to the poor dilution factor,
topology \tpIX has relatively little impact on the
final results. 

\subsection{Additional corrections}
As summarized in Ref.~\cite{eg1cpip}, radiative corrections
were found to be negligible. The correction from the slightly
polarized nitrogen in the ammonia targets was also found to
be negligible. 

\subsection{Systematic uncertainties}
The dominant systematic uncertainty on all the asymmetry
results is an overall scale uncertainty from the
beam and target polarizations. 
The uncertainty in $A_{LL}$ is relatively small (1.4\%)
because $P_BP_T$ was well-measured
using $ep$ elastic scattering. The relative 
uncertainty in $A_{UL}$ is larger (4\%) due to the
uncertainty in $P_B$, from which we obtained $P_T$
by dividing $P_BP_T$ by $P_B$.

The other source of normalization uncertainty
is the dilution factor. As discussed in more detail
in Ref.~\cite{inclprc}, the uncertainties in the target
composition correspond to about a 2.5\% relative uncertainty in the 
amount of background subtraction, which corresponds to
1\% to  1.5\% in the asymmetry results, for the 
missing particle topologies, and less than 0.5\%
for the fully exclusive topology.

Another source of systematic uncertainty is in the 
factor  $R_{A>2}$. We compared
three methods of determining this factor: a study of
inclusive electron scattering rates, fits to the
low electron-pion missing mass spectra, and the value that gives
the best agreement for $A_{LL}$ between the fully
exclusive topology and the topology where the
recoil nucleon is not detected. This last technique
relies on the fact that the fully exclusive topology
has much less nuclear background. From these comparisons,
we estimate a systematic uncertainty of about 2\% (relative)
for $R_{A>2}$. This translates
into approximately 1.5\% (at low $W$) to 2.5\%  (at high $W$)
overall normalization uncertainties on
both $A_{LL}$ and $A_{UL}$.
 
It is also possible for assumptions made in the 
dilution factor fitting, such as the lack of $\phi^*$
dependence, to result in point-to-point systematic
uncertainties. Based on trying out several different functional
forms to the fit, these were found to be much smaller
than the point-to-point statistical uncertainties.
Adding the above sources of uncertainty in quadrature, we 
obtain an overall normalization uncertainty of 3\% for 
$A_{LL}$ and 5\% for $A_{UL}$.

\section{Results}

With over 5700 kinematic points, each with relatively large statistical 
uncertainties, it is a challenge to portray the entire data set in
a meaningful way. For plotting purposes,
we therefore averaged together adjacent bin
triplets or quartets in $W$, and adjacent bin pairs in $Q^2$. 
The complete set of results is available in the
CLAS data base~\cite{clasdb} and in the
Supplemental Material associated with this article~\cite{SMp}.

\subsection{$A_{LL}$}
\begin{figure}[hbt]
\centerline{\includegraphics[width=6.5in,angle=90]{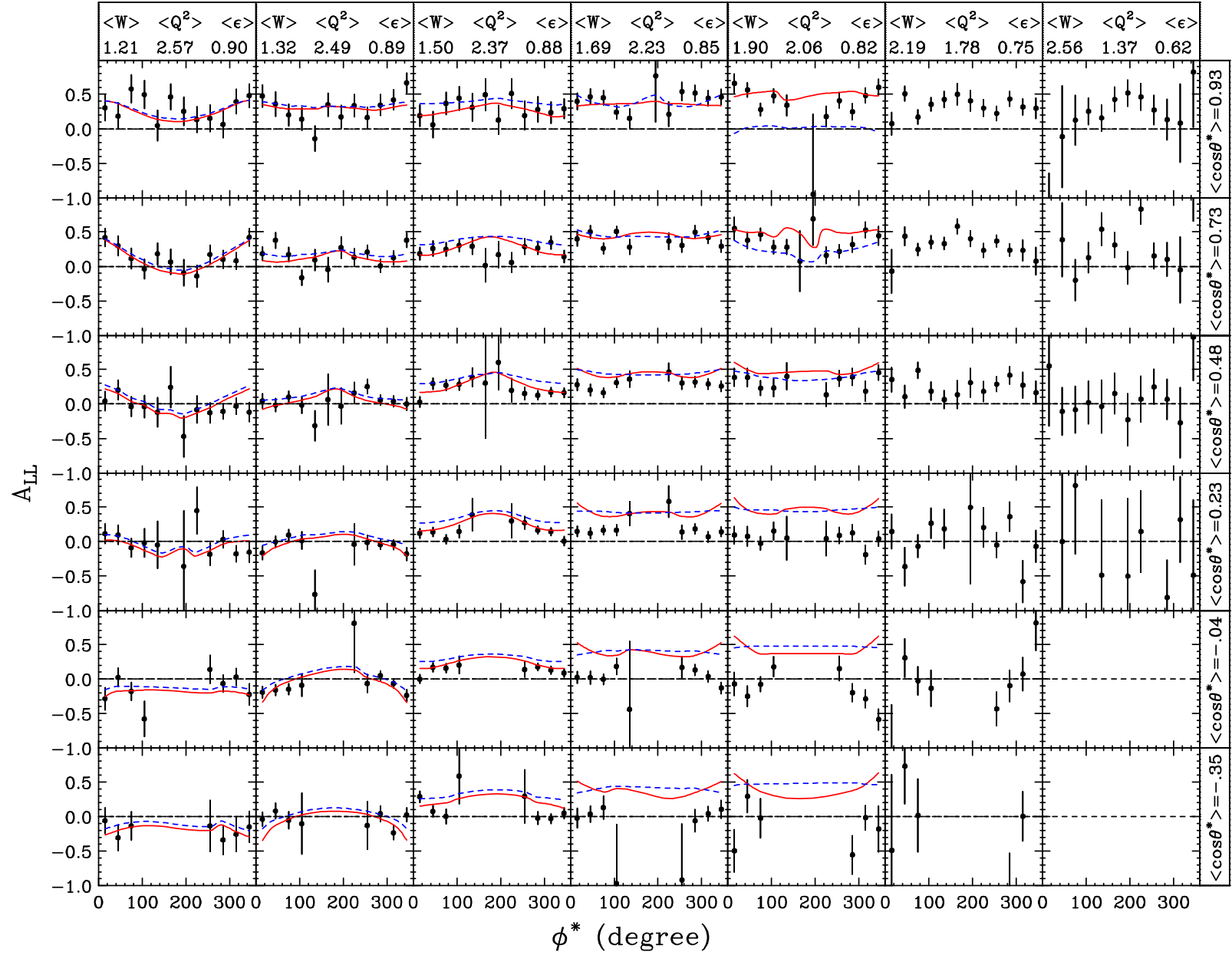}}
\caption{(color online) Beam-target double spin asymmetry $A_{LL}$
for the reaction \tpIIr as a function of  $\phi^*$
in seven bins in $W$ (columns) and 
six \cthcmsp bins (rows). The
results are from the two lower $Q^2$ bins of this analysis.
The error bars reflect statistical uncertainties only.
The solid red curves are from the MAID 2007 fit~\cite{maid} 
and the blue dashed curves are from a JANR fit~\cite{janr}.
}
\label{fig:ALLloq}
\end{figure}

\begin{figure}[hbt]
\centerline{\includegraphics[width=6.5in,angle=90]{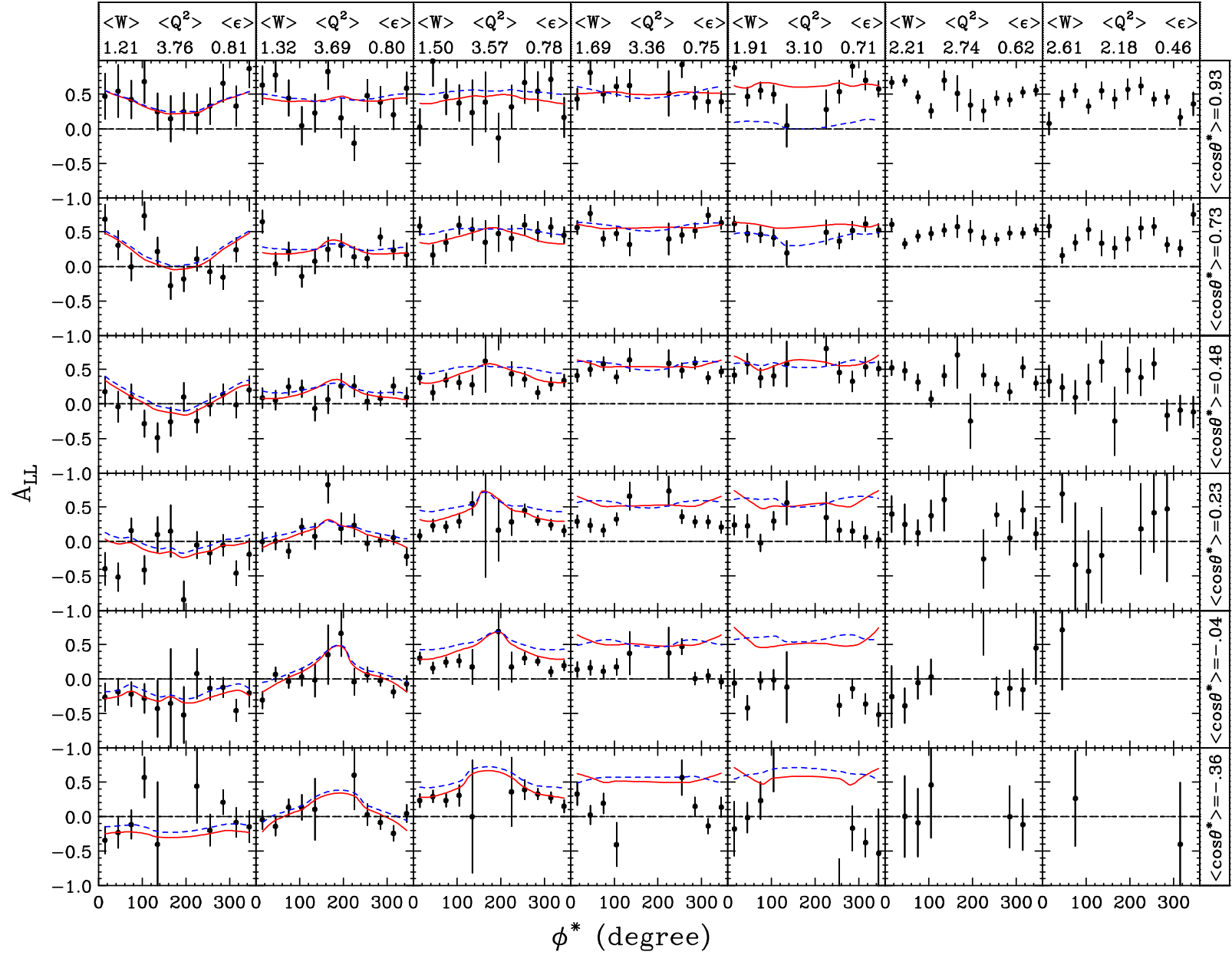}}
\caption{Same as Fig.~\ref{fig:ALLloq}, except for the
 two larger $Q^2$ bins of this analysis.
}
\label{fig:ALLhiq}
\end{figure}

The results for the beam-target spin asymmetry 
$A_{LL}$ are plotted as a function of
$\phi^*$ in seven bins in $W$ and six bins in
$\cos(\theta^*)$ in Fig.~\ref{fig:ALLloq} for the lower
$Q^2$ data and in Fig.~\ref{fig:ALLhiq} for the higher
$Q^2$ data. A weak trend for larger asymmetries at 
larger $Q^2$ can be observed.

The main features of the data is a
relatively large and positive asymmetry (averaging about
0.3) for most kinematic bins. 
A major exception is  
for the lowest $W$ bin, centered on the $\Delta(1232)$
resonance, where the values of $A_{LL}$ are closer to zero.
This feature is expected because the $\Delta(1232)$
transition is dominated by spin-1/2 to 
spin-3/2 transitions, which
gives a negative value of $A_{LL}$, balancing the
positive contribution from the Born terms. 
Another exception is for the lowest \cthcmsp bins,
where again the asymmetries are close to zero. 

Also shown on the plots are the
results of two representative fits to previous data (limited
to $W<2$ GeV):
the 2007 version of the
MAID unitary isobar fit~\cite{maid} and
the Unitary Isobar version of the Joint Analysis
of Nucleon Resonances (JANR) fit~\cite{janr}, averaged with
the same weighting as the data points. Formally, these two
fits are rather similar in nature, but differ in the data
sets used, and in the functional forms used for the 
$Q^2$-dependence of the resonance form factors. 
By and large, both the MAID 2007 and the JANR fits describe
the data reasonably well up to $W=1.6$ GeV,
with  differences appearing at larger $W$. 
Compared to the asymmetries for 
exclusive $\pi^+$ electroproduction from this same
experiment (see figures in Ref.~\cite{eg1cpip}), 
the $\pi^0$ asymmetries are generally closer to zero,
except at forward angles and
larger values of $W$, where they are very similar.


\subsection{$A_{UL}$}
\begin{figure}[hbt]
\centerline{\includegraphics[width=6.5in,angle=90]{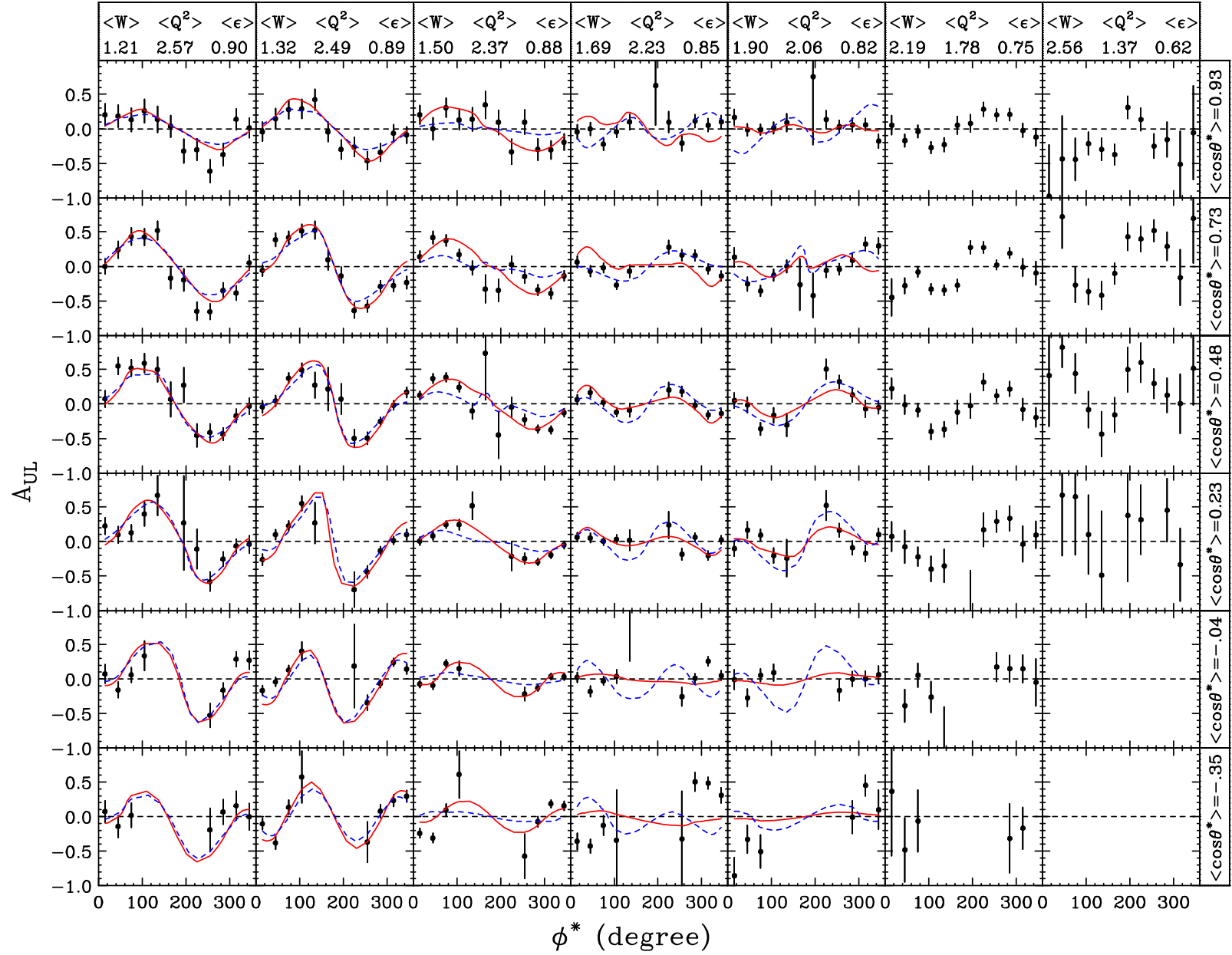}}
\caption{(color online) Target single-spin asymmetry $A_{UL}$
for the reaction \tpIIr as a function of  $\phi^*$
in seven bins in $W$ (columns) and 
six \cthcmsp bins (rows). The
results are from the two lower $Q^2$ bins of this analysis.
The error bars reflect statistical uncertainties only.
The solid red curves are from the MAID 2007 fit~\cite{maid}
and the blue dashed curves are from a JANR fit~\cite{janr}.
}
\label{fig:AULloq}
\end{figure}

\begin{figure}[hbt]
\centerline{\includegraphics[width=6.5in,angle=90]{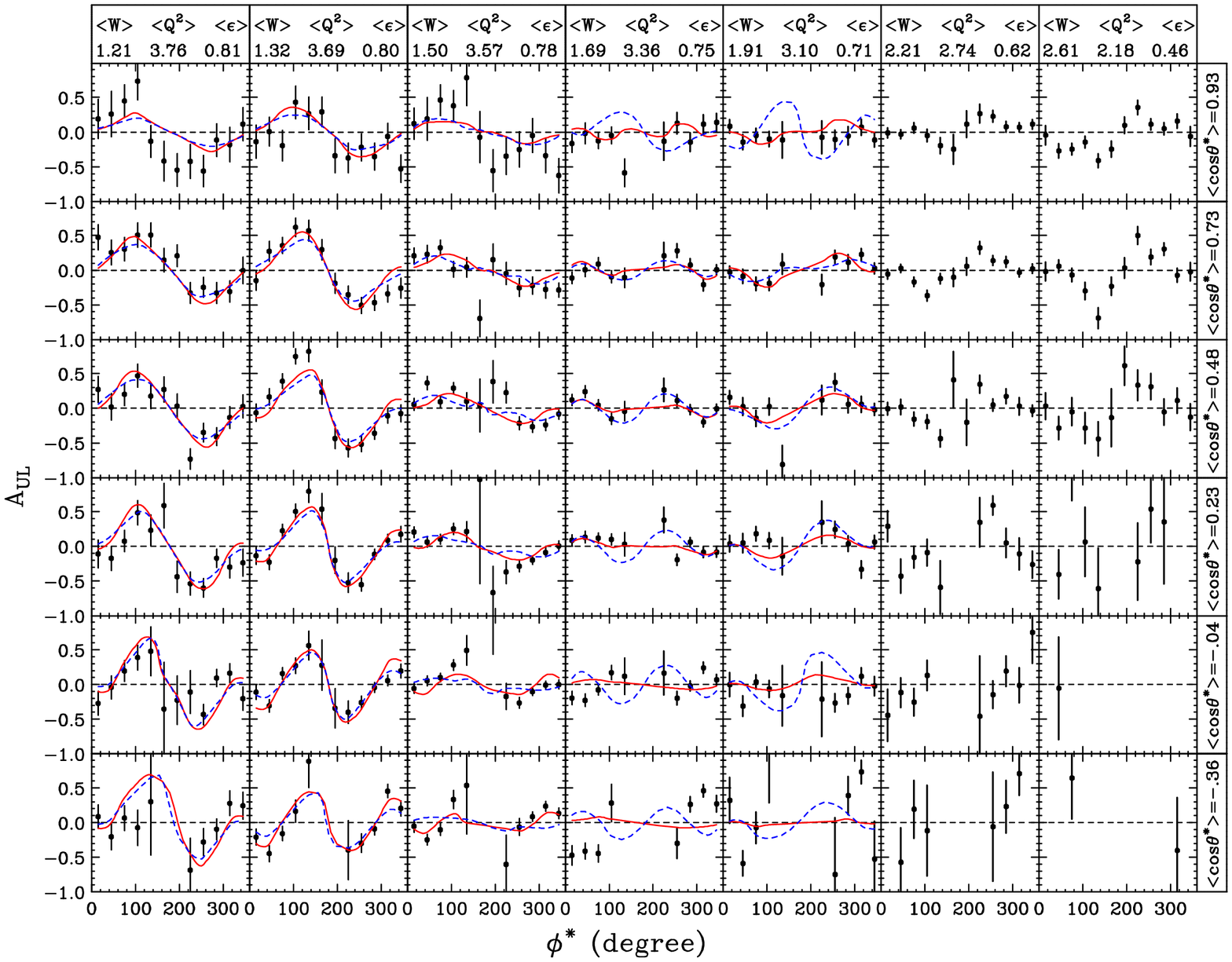}}
\caption{Same as Fig.~\ref{fig:AULloq}, except for the
 two larger $Q^2$ bins of this analysis.
}
\label{fig:AULhiq}
\end{figure}

The results for the target spin asymmetry 
$A_{UL}$ are plotted as a function of
$\phi^*$ in seven bins in $W$ and six bins in
$\cos(\theta^*)$ in Fig.~\ref{fig:AULloq} for the lower
$Q^2$ data and in Fig.~\ref{fig:AULhiq} for the higher
$Q^2$ data. It can be seen that the $Q^2$-dependence
of the results is weak. 
The main feature of the data are a large 
$\sin(\phi^*)$ modulations that are small at forward
angles, and grows to nearly maximal values at
central angles. At low values of $W$, the modulations
are almost equal in magnitude, but of opposite sign,
to those observed for $\pi^+$ electroproduction
(see corresponding figures in Ref.~\cite{eg1cpip}), while
at large values of $W$, the sign of the modulations
changes from the low $W$ asymmetries to be in 
agreement with the $\pi^+$ asymmetries. 

The sign and magnitude of the results
is well reproduced by the MAID and JANR fits
for $W<1.6$ GeV. At larger $W$, the MAID fit reproduces
the relatively small asymmetries observed in the data
for $1.6<W<2$ GeV, while the JANR fit exhibits larger
asymmetries than observed in the experiment. 
Combined with the results
for $A_{LL}$, the results for $A_{UL}$ strongly suggest
that there are important nucleon resonance contributions
to exclusive pion electroproduction for $W>1.7$ GeV
and $Q^2>1$ GeV$^2$. 
 
\section{Summary}
Target and beam-target spin asymmetries in
exclusive  $\pi^0$ electroproduction
($\gamma^* p \to p \pi^0$)
were obtained from scattering of 6~GeV
longitudinally polarized
electrons from longitudinally polarized protons
using the $\rm{CLAS}$ detector at Jefferson Lab. 
The kinematic range covered is $1.1<W<3$ GeV and $1<Q^2<6$
GeV$^2$. Results were obtained for about 5700
bins in $W$, $Q^2$, \cthcm, and $\phi^*$. Except at
forward angles, very
large target-spin asymmetries are observed 
over the entire $W$ region. In contrast to $\pi^+$
electroproduction, the sign of the $A_{UL}$ modulations
changes from positive at low $W$ to negative at
high $W$.  Reasonable agreement is
found with the phenomenological MAID 2007 fit~\cite{maid}
and the JANR fit~\cite{janr} to previous data
for $W<1.6$ GeV, but significant differences are seen
at higher values of $W$, where no data were available when the fits
were made.  We anticipate that new global fits using 
the present $\pi^0$ target and beam-target asymmetry data, when
combined with beam-spin asymmetry and spin-averaged cross
section data, as well as $\pi^+$ observables, 
will yield major insights 
into the structure of the proton and its many excited states.

\section*{Acknowledgments}
We thank I. Aznauryan for providing the JANR source
code and L. Tiator for providing the MAID 2007
source code. 
We thank X. Zheng for suggesting the functional form of
the dilution factor fit.
We acknowledge the outstanding efforts of the staff
of the Accelerator and the Physics Divisions at Jefferson Lab that made
this experiment possible.  This work was supported by
the U.S. Department of Energy (DOE), 
the National Science Foundation,
the Scottish Universities Physics Alliance (SUPA),
the United Kingdom's Science and Technology Facilities Council,
the National Research Foundation of Korea,
the Italian Instituto Nazionale di Fisica Nucleare, the French Centre
National de la Recherche Scientifique, and the French Commissariat \`{a}
l'Energie Atomique.
This material is based upon work supported by the 
U.S. Department of Energy, Office of Science, 
Office of Nuclear Physics under contract 
DE-AC05-06OR23177.

\end{document}